\documentclass[conference]{IEEEtran}
\IEEEoverridecommandlockouts
\usepackage[table]{xcolor}
\usepackage{algorithm}
\usepackage{lipsum}  
\usepackage{cite}
\usepackage{dblfloatfix}
\usepackage{float}
\usepackage{comment}
\usepackage{amsmath,amssymb,amsfonts}
\usepackage{mathtools}
\usepackage{algorithm}
\usepackage{steinmetz}
\usepackage[noend]{algpseudocode}
\usepackage{graphicx}
\graphicspath{{./Figures/}}
\usepackage{textcomp}
\usepackage{siunitx}
\usepackage{url}
\usepackage{array,multirow}
\usepackage[table]{xcolor}
\usepackage{xcolor,colortbl}
\def\BibTeX{{\rm B\kern-.05em{\sc i\kern-.025em b}\kern-.08em
T\kern-.1667em\lower.7ex\hbox{E}\kern-.125emX}}

\begin{document}	
\title{Novel Real-Time EMT-TS Modeling Architecture for Feeder Blackstart Simulations}

\author{\IEEEauthorblockN{Victor Paduani, Bei Xu, David Lubkeman, Ning Lu}
\IEEEauthorblockA{Department of Electrical and Computer Engineering} 
North Carolina State University, Raleigh, NC\\vdaldeg@ncsu.edu
\thanks{This  research  is  supported  by  the  U.S.  Department  of  Energy’s  Office  of Energy  Efficiency  and  Renewable  Energy  (EERE)  under  the  Solar  EnergyTechnologies Office Award Number DE-EE0008770.}
}
\maketitle

\begin{abstract}


\textbf{This paper presents the development and benchmarking of a novel real-time electromagnetic-transient and transient-stability (EMT-TS) modeling architecture for distribution feeder restoration studies. The work presents for the first time in the literature a real-time EMT-TS testbed in which the grid-forming unit is simulated in EMT domain, operating as the slack bus of the phasor domain while including unbalanced voltage conditions. To evaluate the performance and limitations of the proposed model, an equivalent EMT testbed is developed to serve as a benchmark. First, the co-simulation framework and the domain coupling methodology are introduced. Next, the steady-state operation of the EMT-TS and EMT models are matched. Then, tests are conducted to analyze the transient performance of the proposed EMT-TS model when compared to its EMT benchmark. Results reveal that when utilizing a battery energy storage system (BESS) as the grid-forming unit, the EMT-TS testbed can maintain high accuracy for typical load steps.}

\end{abstract}

\begin{IEEEkeywords}
Co-simulation, EMT-TS, hardware-in-the-loop,  real-time simulation, transients.
\end{IEEEkeywords}

\section{Introduction}

As the participation of inverter-based resources (IBRs) in the grid increases, new simulation methods that can integrate the fast response of power electronic devices must be developed \cite{PESmag2021}. Conventional transient stability (TS) software packages that run at the millisecond level are sufficient to simulate synchronous generator dynamic behaviors in a larger network. However, as synchronous generators are being displaced by energy systems powered by IBRs with advanced grid-support functionalities \cite{paduani2020maximum}, \cite{paduani2021implementation}, electromagnetic transient (EMT) simulations are required for capturing IBR dynamic responses at the microsecond level.

EMT tools provide a greater depth of analysis for a wide frequency range, whereas TS is less accurate if there are high frequency dynamics in the grid. Nevertheless, the computational effort associated with EMT simulations make them an unviable solution when simulating realistic distribution networks \cite{jain2021integrated}. Therefore, by integrating EMT-modeled power electronic systems into the TS-modeled power transmission and distribution networks, co-simulations can become a reliable alternative for modeling grids with high penetration of IBRs. As a consequence, the interest in EMT-TS co-simulations have been growing in both industry and academia in recent years \cite{shu2017novel}.

In addition, by developing co-simulation models in real-time simulators, hardware-in-the-loop (HIL) systems can be used to test protection schemes in large networks while including the dynamics of IBRs spread through the grid. Because an HIL testbed can also represent realistic communication links between distributed controllable devices and their centralized controllers, energy management systems (EMS) coordinating the operation of IBRs can be developed and tested in real-time or faster-than-real-time, day-long settings \cite{shirsat2020hierarchical}. 


In \cite{mongrain2019real}, Mongrain \emph{et al.} introduced a real-time simulation test system with established TCP/IP communication links on the eMEGASIM platform from OPAL-RT, but no phasor domain is included. A EMT-TS model built also on eMEGASIM is presented by Athaide \emph{et al.} in \cite{athaide2019matlab}, but it is not a real-time model. Researchers from RTDS Technologies recently presented a model with a real-time EMT-TS coupling via a Dynamic Phasor Line (DPL) \cite{konara2020interfacing}. However, the method requires the DPL to have a propagation delay larger than the EMT domain timestep with only a maximum of four coupling points allowed. Most importantly, there is a need for quantitative studies that can highlight how much information is lost by moving from an EMT model into an EMT-TS model, especially for real-time operation. In \cite{song2020research}, Song \emph{et al.} compared the co-simulation model with the network reduction model and analyzed the computational costs of the two methods. Still, not enough is presented related to the EMT and EMT-TS performance benchmarking.

\setlength{\parskip}{0pt}
To resolve the above issues, in this paper, we present a novel real-time EMT-TS co-simulation modeling architecture, and conduct tests to validate the performance of the proposed testbed against an equivalent model built in EMT domain. The paper presents the following contributions to the literature.

\begin{itemize}
    \item Introduces for the first time in the literature a real-time EMT-TS testbed in which the grid-forming unit is modeled in EMT domain, operating as the slack bus of the phasor domain including unbalanced voltage conditions.
    \item Analyzes the transient performance of the proposed model for different grid operation conditions, and compares the results with an equivalent model developed in EMT domain.
    \item Presents a coupling method for multiple coupling points between the EMT and phasor domains for moving devices across the feeder into the EMT domain.
\end{itemize}

\section{Methodology}

\begin{figure}[htb]
    \centering
    \includegraphics[width=0.5\textwidth
	]{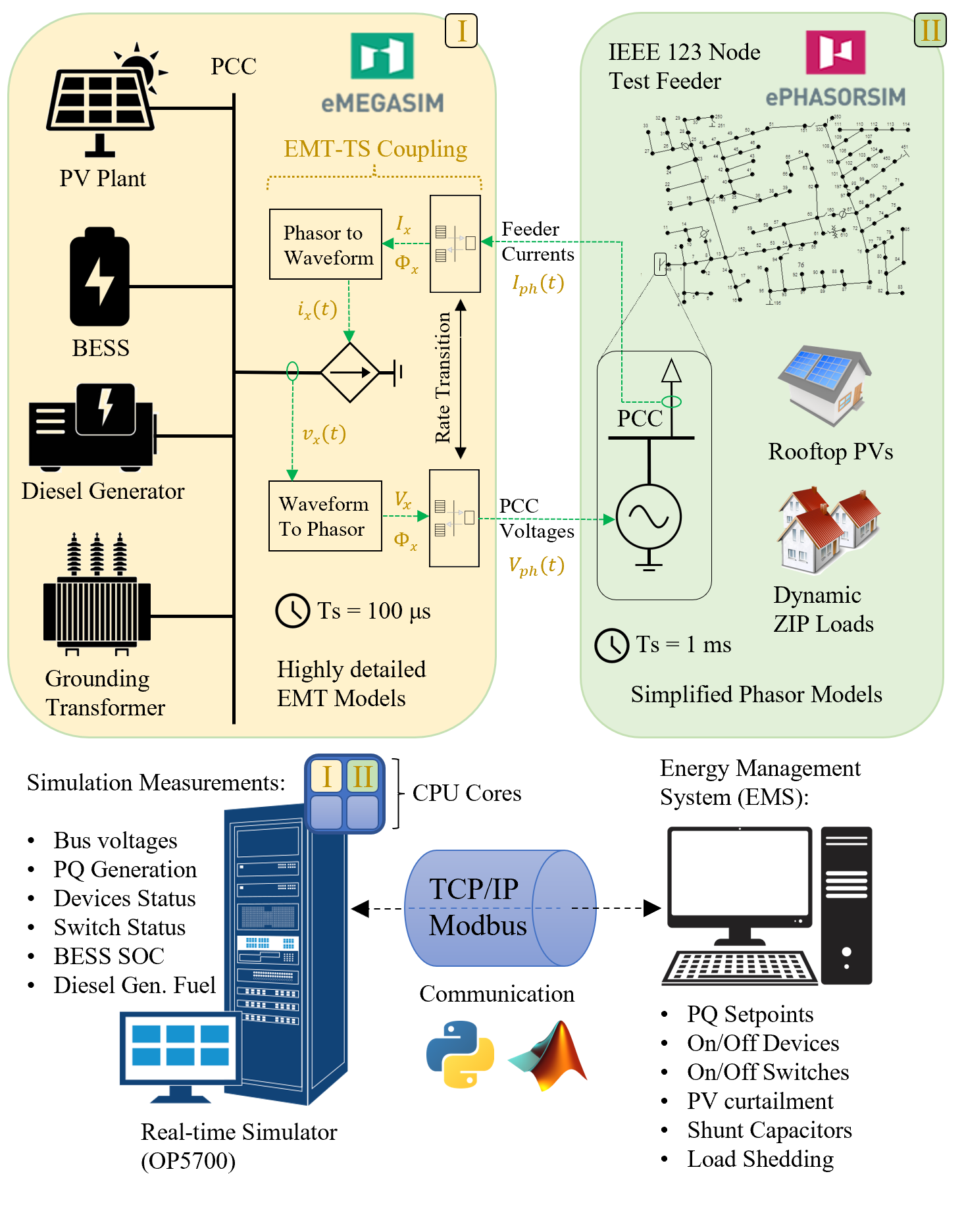}
    \caption{Proposed EMT-TS co-simulation testbed framework.}
    \label{system_fig}
\end{figure}

Figure \ref{system_fig} displays the EMT-TS testbed, which consists of a microgrid connected to a distribution feeder via a point of common coupling (PCC). The components are split into two subsystems. The first subsystem includes distributed energy resources, e.g., a grid-forming battery energy storage system (BESS), utility-scale PV systems, diesel generators (DGs), and the grounding transformer. This subsystem is simulated at the microsecond level in eMEGASIM. The second subsystem includes distributed rooftop PVs, shunt capacitor banks, voltage regulators, ZIP load models, and the unbalanced IEEE 123-bus network model \cite{ieee123} used to emulate the opration of realistic distribution feeders. The second subsystem is simulated in ePHASORSIM at the milisecond level. Note that a subsystem can be assigned to more than one core, if needed, but each subsystem requires at least one core for itself.

The EMS algorithms are running externally and can read measurements, set devices status, and send power setpoint commands using Modbus registers. Note that this is a function that a real-time simulation platform will normally provide for modeling communication links between the real-time simulator and an external computer. 


\subsection{EMT-TS Co-Simulation Coupling}



Since both subsystems are simulated in parallel and exchange information with each other in real-time, there is a delay for events to propagate from one susbystem to another. In this paper, we introduce a coupling method in which the microgrid operates as the slack bus of the system, represented as a voltage source in phasor domain. This allows the simulation of feeder blackstart scenarios, in which a microgrid powered by BESS, utility-scale PVs, and/or DGs can be used for picking up sections of the feeder during system restoration.

\subsubsection{Time Interpolation Coupling}

Time interpolation is a commonly used technique for coupling EMT and phasor domains in co-simulations. The method consists of linearly interpolating both magnitude and phases of the signals coming from phasor domain for each EMT domain timestep \cite{plumier2016co}. Thus, in the proposed EMT-TS testbed, the distribution feeder is represented as a current source, with current magnitude ($I_{x}$) and angle ($\phi_{x}$) calculated from the phasor domain current ($I_{ph}$) as follows.


\begin{equation}
    I_{x}(t+nh) =  \lvert I_{\mathrm{ph}}(t) \rvert + \dfrac{n}{N}\Big( \lvert I_{\mathrm{ph}}(t+H) \rvert - \lvert I_{\mathrm{ph}}(t) \rvert \Big)
\end{equation}

\begin{equation}
    \phi_{x}(t+nh) = \angle I_{\mathrm{ph}}(t) + \dfrac{n}{N}\Big( \angle I_{\mathrm{ph}}(t+H) - \angle I_{\mathrm{ph}}(t) \Big)
\end{equation}
where the index $x$ represents phases ($a/b/c$), $h$ and $H$ correspond to the EMT and phasor timesteps, respectively, $N=H/h$, and $n$ ranges from zero to $N$. Thus, the current waveform for each phase is obtained by (\ref{current_eq}).

\begin{equation}
    i_{x}(t + nh) = \sqrt{2} I_{x}(t + nh) \cos{[\omega_{\mathrm{pcc}}(t+nh) +\phi_{x}(t +nh)]}
    \label{current_eq}
\end{equation}
where $\omega_{\mathrm{pcc}}$ is the angular frequency measured at the PCC with a phase-locked loop.

As will be shown in the simulation results section, the time interpolation coupling can smoothen the impact of fast current transients applied in the EMT domain, causing a larger deviation for the response between the EMT and EMT-TS testbeds. Therefore, instead of time interpolation, in this work, we directly update the waveforms with signals coming from the phasor domain with (\ref{ia_waveform}). In that case, to avoid numerical issues in the solver, a first-order low-pass filter with time constant equal to the EMT timestep ($T_{\mathrm{s,emt}}$) is applied on the recreated current signals. Moreover, due to the delay introduced by the filter, a feedforward compensation ($T_{\mathrm{s,emt}}$) is added to (\ref{ia_waveform}).

\begin{equation}
    i_{x}(t) = \sqrt{2}I_{x} \cos{\big[\omega_{\mathrm{pcc}}(t + T_{\mathrm{s,emt}}) + \phi_{x}\big]}
    \label{ia_waveform}
\end{equation}

\subsubsection{Phasor Extraction}
Phasor domain signals are obtained from the three-phase voltages measured at the PCC at a rate given by the phasor domain timestep ($T_{\mathrm{s,ph}}$). The voltages magnitudes are obtained with the true root mean square (RMS) value of the corresponding waveform,  whereas the angles are obtained with its fundamental value, which can be obtained with Fourier analysis (\ref{fundamental_angle}). Both the RMS and phase calculations are performed over a running average window of one cycle of the signal's fundamental period ($T$).

\begin{equation}
    a = \frac{2}{T} \int_{(t-T)}^{t} v(t)\cos(\omega_{0}t) \,dt 
\end{equation}

\begin{equation}
    b = \frac{2}{T} \int_{(t-T)}^{t} v(t)\sin(\omega_{0}t) \,dt 
\end{equation}

\begin{equation}
    \phi_{\mathrm{ph}} = a\mathrm{tan}{(b/a)}
    \label{fundamental_angle}
\end{equation}
 
 \subsection{EMT Domain Benchmark Model}
\begin{figure}[ht]
    \centering
    \includegraphics[width=0.5\textwidth]{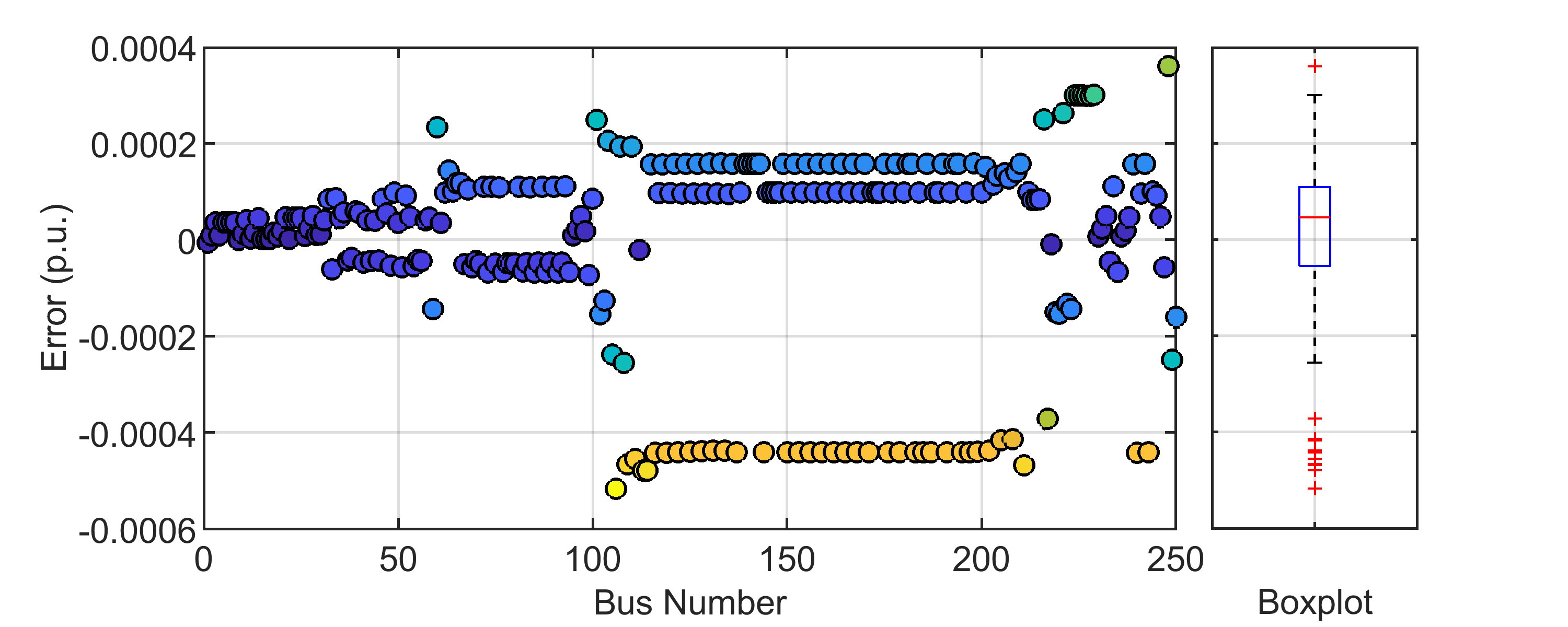}
    \caption{A comparison of the nodal voltage magnitude between the full EMT and EMT-TS testbeds.}
    \label{benchmark}
\end{figure}

To benchmark the EMT-TS testbed performance, an equivalent EMT model with the same network parameters and loading conditions is built using the combined state-space nodal (SSN) modeling method introduced in \cite{dufour2010combined}. 
SSN divides the state-space matrix utilized by Simulink's solver into separate matrices based on nodes defined by the user. 
As shown in Fig. \ref{benchmark}, the nodal voltage error distribution between the EMT and EMT-TS testbeds is negligible in steady state operation with 
the maximum error below 0.0006 p.u. and the interquartile range within $\pm$0.0003 p.u.. In addition, the active and reactive power flow measured at the PCC (EMT-domain side) between both testbeds has the highest mismatch of 1.0 kW (on phase $b$) and 2.6 kVAR (on phase $a$) when supplying peak load. 

\section{Simulation Results}

In this section, we present the following simulation results: (A) transient performance evaluation during load steps, (B) the impact of the TS domain timestep selection on propagation delays, (C) the impact of the time interpolation coupling, (D) a capacitor bank switching event, and (E) computational cost analysis.


\subsection{Transient Performance During Load Steps}
\begin{figure}[t]
    \centering
    \includegraphics[width=0.5\textwidth
	]{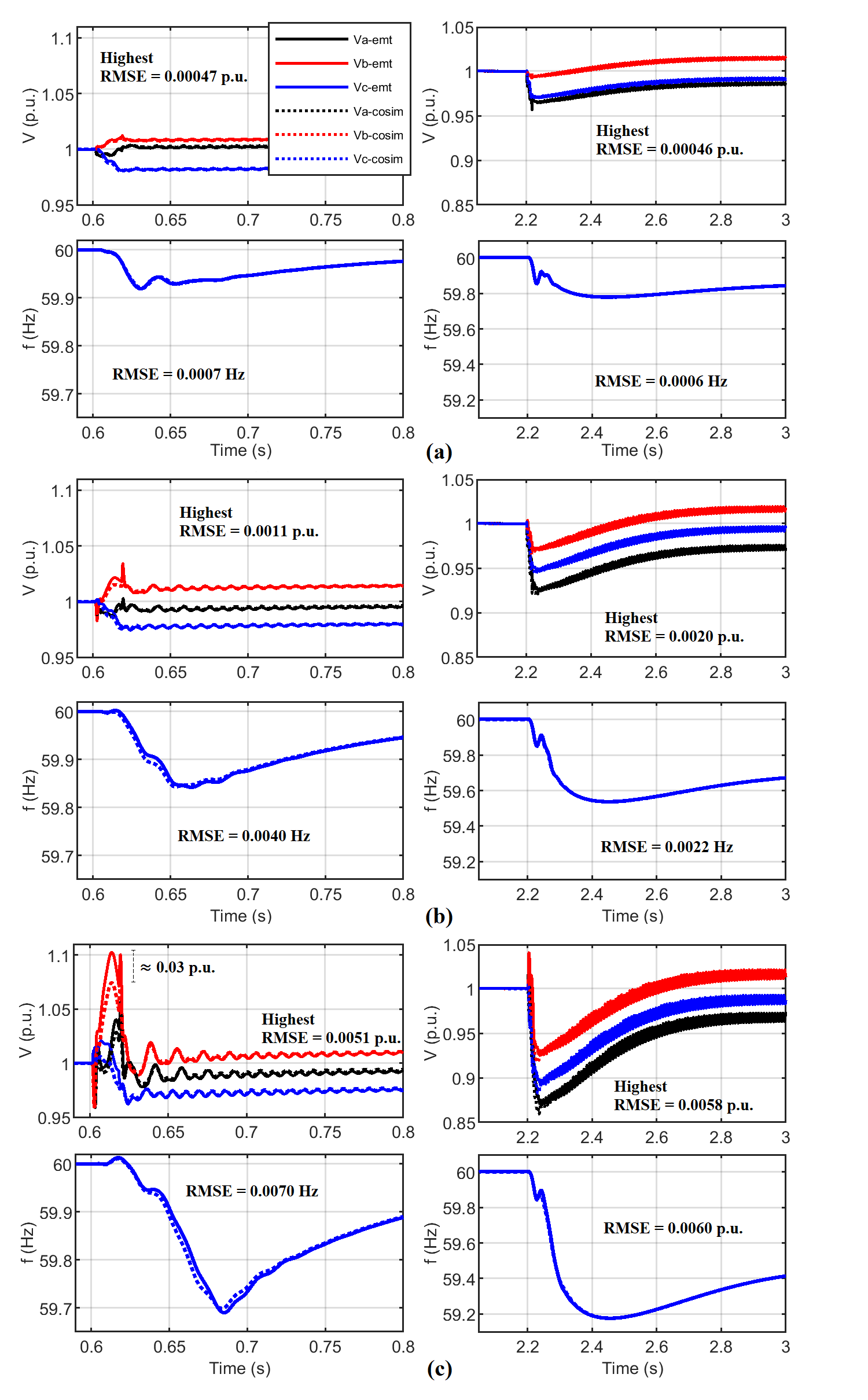}
    \caption{Transient response comparison between the EMT-TS and the EMT testbeds. Plots on the left: 2-MVA BESS. Plots on the right: 3.125-MVA diesel generator. (a) Load step change I; (b) Load step change II, and (c) Load step change III. Refer to Table \ref{load_steps} for the test setup.}
    \label{steps123}
\end{figure}

The load steps applied during testing are presented in Table \ref{load_steps}. Two different types of grid-forming units are analyzed: a 2-MVA BESS (parameters from \cite{xu2021Novel}) and a 3.125-MVA diesel generator (parameters from \cite{unified2021paduani}). 
The voltage and frequency transient responses of the proposed EMT-TS and the full EMT model are shown in Fig. \ref{steps123}. Note that a delay of 2 ms is added to the load steps in the EMT model so that the disturbances are applied at the same point-on-wave (POW) for both testbeds. The RMS error (RMSE) between EMT and EMT-TS curves is included in the figure of each test, with the highest RMSE value corresponding to the phase with the largest deviation.

Because the response of the DG is above the millisecond level, the modeling error of the EMT-TS testbed is negligible. For the BESS unit, the transient error is also negligible in the first load step-change (approximately 25\% of the BESS rated power). In the second load step-change (approximately 50\% of the BESS rated power), a maximum error of 0.006 p.u. is observed. During the third load step change (close to the BESS rated power), a maximum RMS voltage error of 0.03 p.u. is observed. Note that this is an extreme scenario and thus, represents the upper bound of the modeling error. 
It is worth mentioning that for the third load step-change in the diesel generator scenarios, (Fig. \ref{steps123}(c)), phase b (in red) presents a voltage rise at the beginning of the transient. This is because the large, unbalanced load step causes a numerical issue in the DG model. This can be fixed by changing the synchronous machine solver from trapezoidal non-iterative to backward Euler robust (or trapezoidal robust). However, trapezoidal non-iterative was the only solver compatible with the SSN modeling (from the EMT benchmark). Therefore, the solver is kept the same to ensure a consistent comparison between the two testbeds.

\begingroup
\renewcommand{\arraystretch}{1.1}
\begin{table}
    \centering
    \caption{The Setup of Load Step-change Tests}
    \label{load_steps}
    \begin{tabular}{|c|c|c|c|c||c|c|c|}
        \cline{3-8}
        \multicolumn{2}{c|}{}& \multicolumn{3}{c||}{P (kW)}& \multicolumn{3}{c|}{Q (kVAr)}\\\hline
        Step & S (kVA) & A & B & C & A & B & C\\\hline
        I&  448 &196 & 39  & 166 & 97 & 20 & 81 \\\hline
        II & 937 &379 & 174 & 285 & 191 & 85 & 141\\\hline
        III & 1978 & 734 & 490 & 640 & 378 & 243 & 324\\\hline
    \end{tabular}
\end{table}
\endgroup

\subsection{Influence of Coupling Delay}

\begin{figure}[htb]
    \centering
    \includegraphics[width=0.5\textwidth
	]{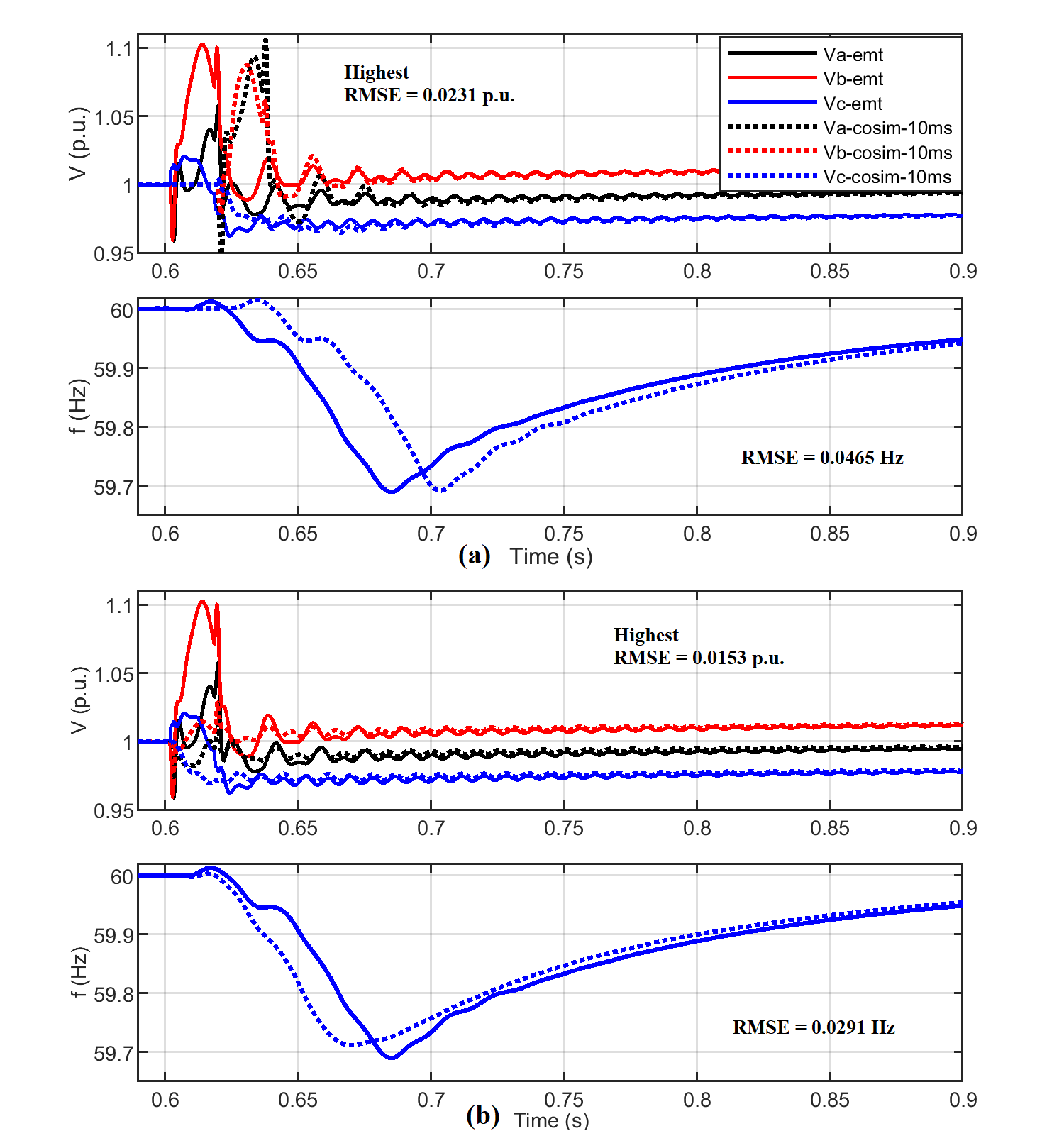}
    \caption{Voltage and frequency transient responses during Step III from Table \ref{load_steps} due to: (a) increasing the communication delay between EMT and phasor domains; (b) utilizing the time interpolation coupling technique}
    \label{delayinterp}
\end{figure}

In this test, the timestep of the phasor domain solver is increased from 1 to 10 ms. As discussed in \cite{xie2019asynchronous}, the propagation delay of events between the EMT and TS domains can be as high as (2T$_{\mathrm{s,ph}}$+ T$_{\mathrm{s,emt}})$. Hence, the timestep change increases the maximum coupling delay between the EMT and co-simulation testbeds from 2.1 to 20.1 ms. The delayed transient can be observed in Fig. \ref{delayinterp}(a). Furthermore, since the transient is affected by the POW when the load-step is applied, the voltage mismatch between the EMT and co-simulation testbeds will increase accordingly.  

\begin{figure}[htb]
    \centering
    \includegraphics[width=0.5\textwidth
	]{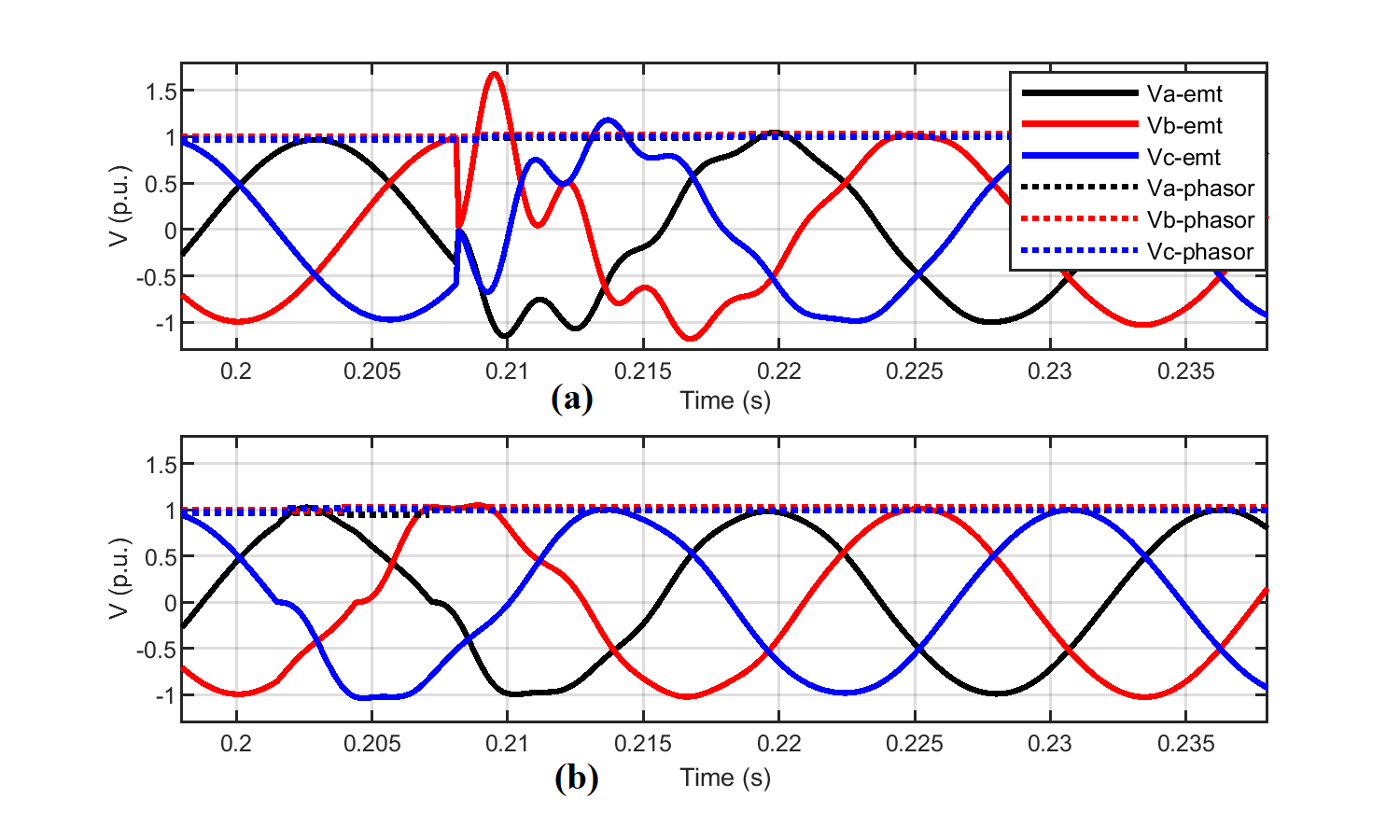}
    \caption{Capacitor bank switching event: (a) peak-voltage energization (worst-case scenario); (b) Zero-voltage-crossing energization.}
    \label{calswitch0}
\end{figure}

\subsection{Influence of Interpolation}

Figure \ref{delayinterp}(b) displays the transient response of the co-simulation testbed during the third load step when including the interpolation method described in Section II-C. Clearly, the interpolation considerably smoothed the response, causing a significant reduction of both voltage and frequency transients. This is an important finding that must be considered when developing a co-simulation testbed in which the grid-forming unit (or slack bus) is in EMT domain.

\subsection{Capacitor Switching Transients}

A capacitor bank is turned on at bus 83 from the feeder. In EMT domain, the bank is modeled as ideal Yg capacitances, whereas in the EMT-TS testbed it is modeled inside ePHASORSIM as shunt loads. Figure \ref{calswitch0} displays the voltages at bus 83 when the capacitors are switches at (a) peak grid voltage (worst-case scenario), and (b) zero-voltage crossing (ZVC).

Evidently, the co-simulation does not capture the peak voltage of 1.6 p.u. observed in the EMT model in case (a). In fact, there is no significant difference between (a) or (b) for the EMT-TS model, since the phasor domain solver does not simulate the capacitors' differential equations. Thus, as long as the capacitor bank of the system in study contains protective measures such as ZVC, the co-simulation can be utilized. Next, we present a method to further improve the performance of the co-simulation for case (b).

\begin{figure}[htb]
    \centering
    \includegraphics[width=0.5\textwidth
	]{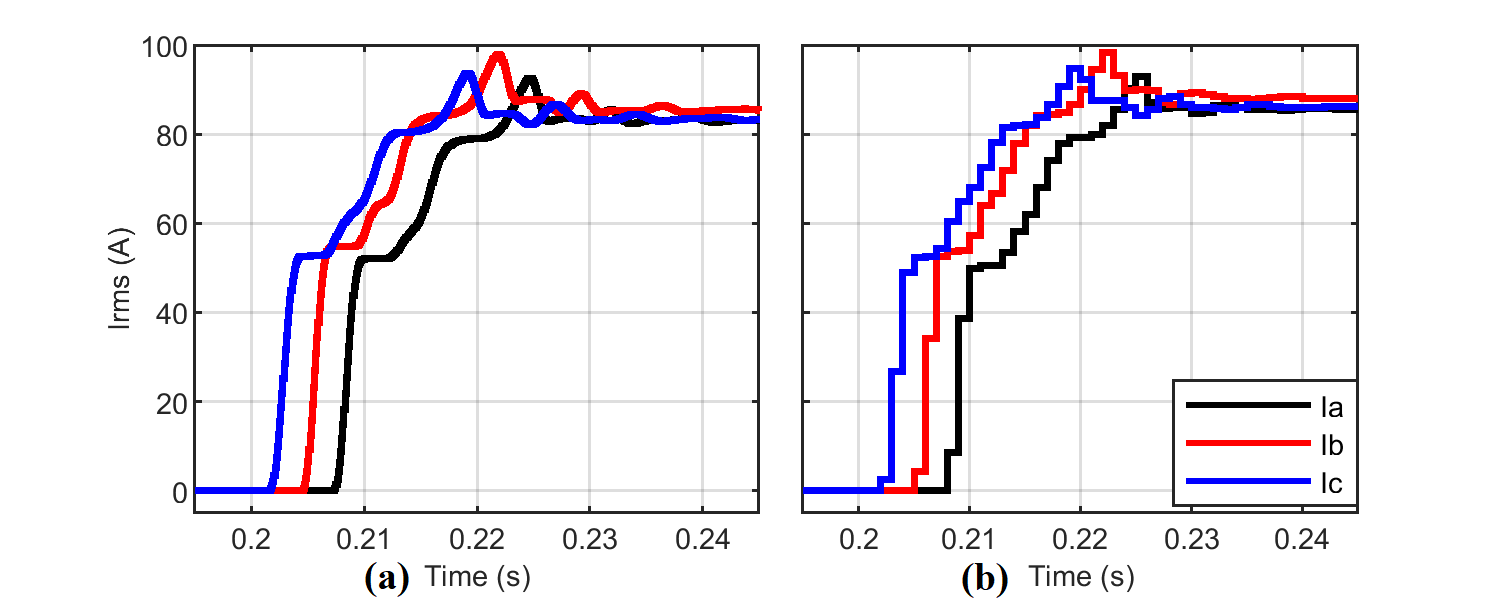}
    \caption{Current injection of the capacitor bank: (a) EMT testbed; (b) EMT-TS testbed with the shunt capacitors modeled in EMT domain.}
    \label{calswitch1}
\end{figure}

\begin{figure}[htb]
    \centering
    \includegraphics[width=0.5\textwidth
	]{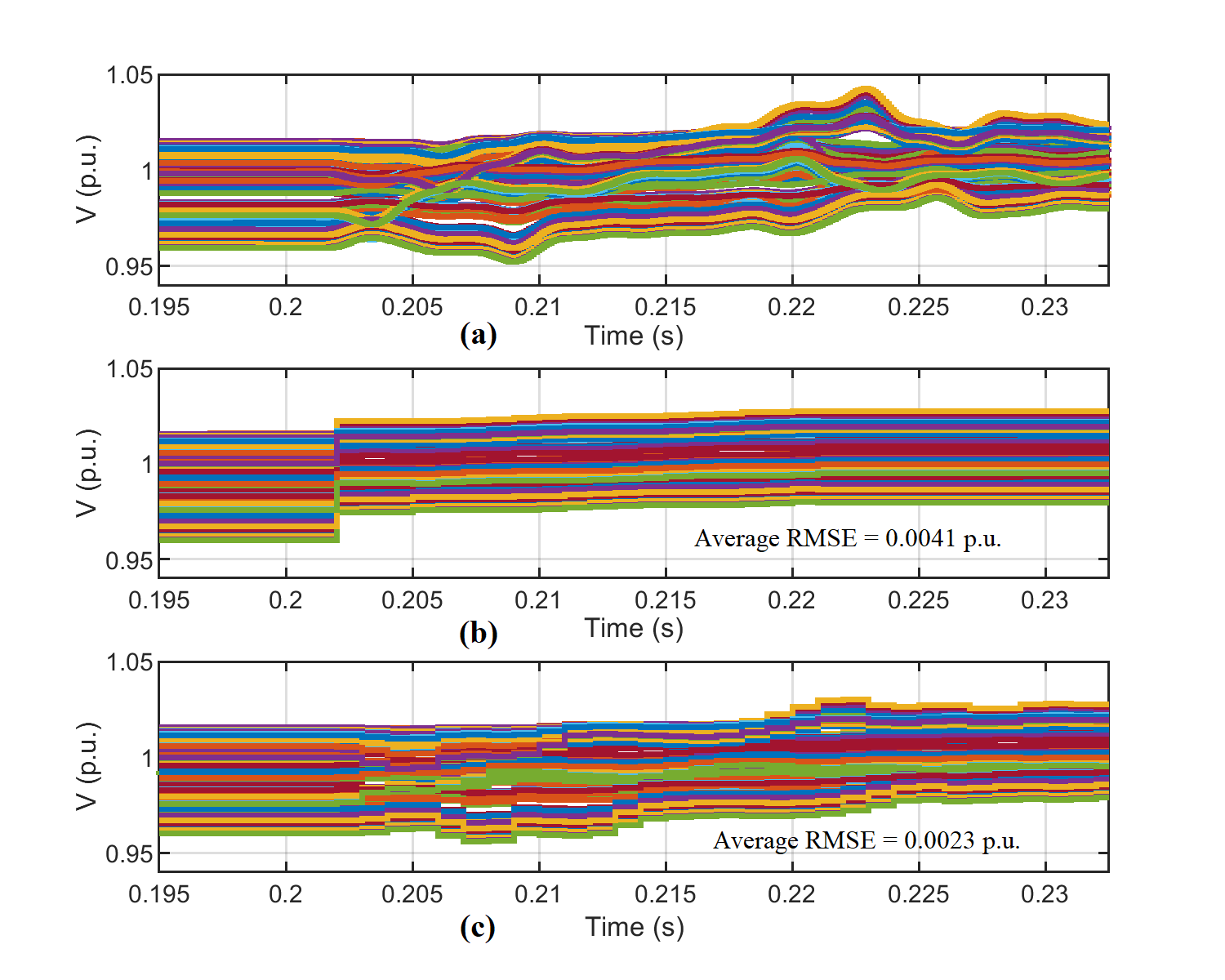}
    \caption{Feeder bus voltages during capacitor bank switching: (a) EMT model; (b) EMT-TS model with shunt capacitors modeled in phasor domain; (c) EMT-TS model with shunt capacitors modeled in EMT domain.}
    \label{calswitch2}
\end{figure}

By adding a second EMT-TS coupling (at bus 83), the capacitors can be modeled in the EMT domain. First, the equivalent impedance between the feeder head and bus 83 is found, and a series RLC circuit is built. Then, the domains are coupled by (i) inserting a voltage source in EMT domain based on voltage phasors measured at bus 83, and (ii) inserting a current source to bus 83 of the phasor domain corresponding to the capacitors' currents in EMT domain. Figure \ref{calswitch1} displays a comparison between the RMS currents measured for the EMT and EMT-TS testbeds for case (b), whereas Fig. \ref{calswitch2} compares the feeder bus voltages during the switching event. Clearly, when simulating the capacitors in EMT domain, their impact on the feeder bus voltages can be better captured. This demonstrates the possibility of utilizing multi-EMT-TS couplings to move devices distributed across the feeder into the EMT domain as needed.

\subsection{Computational Cost Analysis}

The simulator used is OPAL-RT OP5700. The EMT testbed is simulated on one core whereas the EMT-TS testbed is modeled on two cores (minimum required). For a fair comparison, we set the EMT model to run four times slower than real-time in one core, and the EMT-TS testbed to run two times slower than real-time in two cores. The EMT model required an average of \SI{295.4}{\micro\s} to process each \SI{100}{\micro\s} step, consuming approximately 73\% of one core when running four times slower than real-time. The EMT-TS testbed required an average of \SI{21.2}{\micro\s} per \SI{100}{\micro\s} step, consuming around 11.4\% of two cores when running two times slower than real-time.

\section{Conclusion}

A novel real-time EMT-TS co-simulation testbed is presented, in which a grid-forming unit modeled in EMT domain operates as a slack bus in the phasor domain. The proposed testbed is validated against an equivalent EMT model developed as a benchmark. Results demonstrate that the proposed model can capture the transient response of BESS units for typical load steps. Furthermore, we demonstrate that (i) the major impact of increasing the phasor domain timestep is the POW time-drift due to the propagation delay; (ii) using EMT-TS time interpolation coupling method will cause a smoothing effect when modeling fast changing waveforms, and (iii) the framework can be used for multi-EMT-TS coupling so that devices can be moved to the EMT domain such that their transient responses can be captured in higher detail.



\bibliographystyle{IEEEtran}
\bibliography{mybibtex}

\end{document}